\documentclass[12pt]{article}
\begin{document}
\title{Elliptic and hyperelliptic magnetohydrodynamic equilibria}
\author{H. Tasso\footnote{het@ipp.mpg.de}, G. N.
Throumoulopoulos\footnote{gthroum@cc.uoi.gr} \\
$^\star$
Max-Planck-Institut f\"{u}r Plasmaphysik \\
Euratom Association \\  85748 Garching bei M\"{u}nchen, Germany \\
$^\dag$ University of Ioannina,\\ Association Euratom-Hellenic
Republic,\\ Department of Physics, GR 451 10 Ioannina, Greece}
\maketitle
\newpage
\begin{abstract}
 The present study is a continuation of a previous one
 on "hyperelliptic" axisymmetric equilibria started in
[Tasso
and Throumoulopoulos, Phys. Plasmas 5, 2378 (1998)].
  Specifically,
some
equilibria with incompressible flow nonaligned with the magnetic field and
restricted by appropriate side conditions like "isothermal" magnetic surfaces,
"isodynamicity" or $P + B^{2}/2$ constant on magnetic surfaces are found to be
reducible to elliptic integrals. The third class  recovers recent equilibria
found in [Schief, Phys. Plasmas 10, 2677 (2003)]. In contrast
to field aligned flows, all solutions found here have nonzero toroidal magnetic
field on  and elliptic surfaces   near the magnetic axis.
\end{abstract}

PACS: 52.30.Bt, 47.65.+a, 02.30.Jr

\newpage

\section{Introduction and basic equation}

A generalized Grad-Shafranov equation has been derived in Ref. \cite{tas1}
 (Eq.  (22) therein)
to describe
axisymmetric magnetohydrodynamic equilibria with incompressible flows.
 This
equation consisting   the starting point of the present
investigation is given by
\begin{eqnarray}
(1 - M^{2}) \Delta^{*} \psi -\frac{1}{2}(M^{2})'|{\bf\nabla}\psi|^{2} +
\frac{1}{2} (\frac{X^{2}}{1 - M^{2}})'\nonumber \\+ R^{2} (P_{S}(\psi) -
\frac{XF'\Phi'}{1 - M^{2}})' + \frac{R^{4}}{2} (\frac{\rho (\Phi')^{2}}{1 -
M^{2}})' = 0
\end{eqnarray}
along with a Bernoulli relation for the pressure,
\begin{equation}
P = P_{S}(\psi) - \rho [\frac{v^{2}}{2} + \frac{\Phi'\Theta}{\rho}],
\end{equation}
where $P_{S}(\psi)$ is part of the pressure which depends on
$\psi$ only,  $\psi$ being the poloidal magnetic flux function.
The elliptic operator $\Delta^{*}$ is defined by $\Delta^{*} =
R^{2}{\bf \nabla} \cdot ({\bf \nabla }/R^{2})$, $M^{2} =
(F'(\psi))^{2}/\rho$ where $F(\psi)$ is the poloidal stream
function and $\rho (\psi)$ is the mass density, $\Phi (\psi)$ is
the electrostatic potential, $\Theta/(\rho R)$  is the toroidal
velocity component  and $X(\psi) = I(1 - M^{2}) + R^{2}F'\Phi'$
where $I/R$ is the toroidal magnetic field. $R, \phi, z$ are the
usual cylindrical coordinates with $z$ corresponding to the axis
of symmetry. As stated in Ref. \cite{tas1} the surface quantities
$F(\psi), \Phi (\psi), X(\psi), \rho (\psi)$ and $P_{S}(\psi)$ are
free functions. For each choice of this set of five functions,
Eq.(1) is fully determined and can be solved whence the boundary
condition for $\psi$ is given.

For our further investigation it is convenient to simplify Eq.(1)
by introducing the following transformation

 \begin{equation}
u(\psi) = \int_{0}^{\psi}[1 - M^{2}(g)]^{1/2} dg,
\end{equation}
which reduces (1) to
\begin{eqnarray}
\Delta^{*}u + \frac{1}{2} (\frac{X^{2}}{1 - M^{2}})'\nonumber \\+ R^{2}
(P_{S}(u) - \frac{X\Phi'F'}{1-M^{2}})' + \frac{R^{4}}{2} (\frac{\rho
(\Phi')^{2}}{1 - M^{2}})' = 0,
\end{eqnarray}
where the primes indicate now the derivatives with respect to $u$
but $F^\prime= d F/d\psi$ and $\Phi^\prime= d \Phi/d\psi$. (See
previous work e.g. Ref. \cite{thr}).  Note that no quadratic term
in $|{\bf\nabla}u|^{2}$ appears anymore in Eq.(4).

 The paper is organized as
follows: section 2 addresses the question of the side conditions
while in section 3 the shape of the magnetic surfaces is
determined near magnetic axis. The conclusions are in section 4.

\section{Side conditions}

 Instead of specifying the free functions mentioned above to
determine Eq.(1), it may be of physical or mathematical importance
to introduce side conditions on some physical quantities like the
total pressure, the magnitude of the magnetic field or
combinations of them. It is indeed plausible to assume isothermal
magnetic surfaces in hot plasmas \cite{tas1} because of the huge
parallel heat conductivity or to try to eliminate neoclassical
effects \cite{pal} through an isodynamic condition. Such side
conditions lead, in general, to an additional relation between
$(\nabla u)^{2}, u $ and $R$ as already accomplished in section 4
of Ref. \cite{tas1} or in Ref. \cite{pal}. It turns out that, due
to the assumed incompressibility of the flow, Eq.(4) as well as
the side conditions considered here have quartic $R$ dependence on
the right hand side, which together with Eq.(4) can be expressed
as follows

\begin{eqnarray}
\Delta^{*}u = - f(u) - R^{2}g(u) - R^{4}h(u), \\
|{\bf\nabla}u|^{2} = 2[i(u) + R^{2}j(u) + R^{4}k(u)],
\end{eqnarray}
where
\begin{eqnarray}
f(u) = \frac{X^{2}}{2(1 - M^{2})},\\
g(u) = (P_{S} - \frac{X\Phi'F'}{1 - M^{2}})',\\
h(u) = (\frac{\rho\Phi'^{2}}{2(1 - M^{2})})',
\end{eqnarray}
 and the other coefficients depend upon the specific side
condition chosen. Let us consider first the case of isothermal
magnetic surfaces already treated in Ref. \cite{tas1} in the
variable $\psi$. Note that our present variable u defined in (3)
is a special relabeling of the variable $\psi$.

\subsection{Isothermal magnetic surfaces}

 Setting the plasma pressure $P$ as a function of $u$ and using Ref.
\cite{tas1} to calculate the coefficients entering Eq.(6), one
obtains

\begin{eqnarray}
i(u) = - \frac{X^{2}}{2(1 - M^{2})},\\
j(u) = (1 - M^{2})[\frac{P_{S} - P}{M^{2}} - \frac{X\Phi'F'}{(1 -
M^{2})^{2}}],\\
k(u) = \frac{\rho\Phi'^{2}}{2(1 - M^{2})}(\frac{1 - 2M^{2}}{M^{2}}).
\end{eqnarray}
To solve equations (5) and (6) simultaneously we use the method of
section 4 of Ref. \cite{tas1} which boils down to an ordinary
differential equation on each magnetic surface

\begin{equation}
  \left.\frac{\partial z}{\partial x}\right|_{u}=
 -\frac{p}{q} = \frac{\pm\frac{1}{4}[(g+j^\prime)x + \frac{1}{2}(h +k^\prime)x^2
-d]}
 {\left\{2(i+jx + kx^2)-\frac{x}{4}\left\lbrack(g+j^\prime)x +
       \frac{1}{2}(h+k^\prime)x^2 -d\right\rbrack^2\right\}^{1/2}},
\end{equation}
where $x$ stays for $R^{2}$ and five compatibility conditions for
seven free functions including the five free functions of Eq. (1).
There should be no problem, in general, to satisfy those
compatibility conditions. The solutions of (13) are, in general,
hyperelliptic integrals \cite{byr}, which are not related to known
special functions unless they can be reduced to elliptic
integrals. This occurs, in particular, for field aligned flows
($\Phi' = 0$) considered in Ref. \cite{tas1,tas2}. The purpose of
this contribution is to find other cases of elliptic reduction
with nonaligned flows ($\Phi' \neq 0$). The easiest case of that
kind is to annihilate the coefficient of the largest powers of $x$
in the denominator of Eq.(13)

\begin{equation}
h + k' = 0
\end{equation}
 with $h$ and $k$ both different from zero.

 After introducing $J = \frac{\rho\Phi'^{2}}{1 -
M^{2}}$ and using (9) and (12), Eq. (14) leads to
\begin{equation}
J' - \frac{(M^{2})'}{M^{2}(1 - M^{2})}J = 0,
\end{equation}
whose general solution is
\begin{equation}
J = C\frac{M^{2}}{1 - M^{2}},
\end{equation}
with $C \geq 0$ and $0 \leq M^{2} < 1$. For $C = 0$, we recover the case of
field aligned flows already obtained in Ref. \cite{tas1}.

 As a byproduct of this
investigation misprints have been found in the  nonumbered
equations for $k(\psi)$ and $g(\psi)$  after Eq.(35) of Ref.
\cite{tas1}.  They should be corrected as follows:
 \begin{equation}
 k(\psi)= \frac{1}{2}[\frac{\rho(\Phi^\prime)^2}{M^2}-
          \frac{(F^\prime \Phi^\prime)^2}{(1-M^2)^2}] \nonumber
 \end{equation}
 and
 \begin{equation}
g(\psi)= \frac{M^2}{1-M^2}(\frac {P_s}{M^2})^\prime
 -[\frac{X\Phi^\prime F^\prime}{(1-M^2)^2}] +
\frac{(M^2)^\prime}{M^2(1-M^2)}P.\nonumber
 \end{equation}

\subsection{Isodynamic field or $B^{2}$ = function of $u$}

The setting of $B^{2}$ as a function of $u$ (Ref. \cite{tas1} is used for the
calculations)
leads to the functions $i(u)$, $j(u)$ and
$k(u)$ entering the side condition (6) as
\begin{eqnarray}
i(u) = - \frac{X^{2}}{2(1 - M^{2})},\\
j(u) = (1 - M^{2})[\frac{B^{2}}{2} + \frac{X\Phi'F'}{(1 - M^{2})^{2}}],\\
k(u) = - \frac{\rho\Phi'^{2}}{2(1 - M^{2})}M^{2},
\end{eqnarray}
while $f(u)$, $g(u)$ and $h(u)$ are still given by (7)-(9) since Eq.  (5)
 does not
change. Now the reduction equation (14) becomes
\begin{equation}
J' - \frac{(M^{2})'}{1 - M^{2}}J = 0,
\end{equation}
whose solution is
\begin{equation}
J = \frac{C}{1 - M^{2}},
\end{equation}
with $C \geq 0$ and $0 \leq M^{2} < 1$. Again we recover for $C = 0$ the Palumbo
solution for field aligned flows  \cite{tas1,tas2,pal}.

\subsection{($P + B^{2}/2$) = function of $u$}

Though this side condition is not as relevant to hot plasmas as
the previous cases, it is of mathematical interest since it
induces a "hidden symmetry" in the equilibrium equations as
discovered in Ref. \cite{sch}, which leads to a rich class of
solutions. After calculating $(P + B^{2}/2)$ and setting it a
function of $u$, we obtain the coefficients $i(u)$, $j(u)$ and
$k(u)$ entering condition (6) as
\begin{eqnarray}
i(u) = - \frac{X^{2}}{2(1 - M^{2})},\\
j(u) = 2[P - P_{S} + \frac{B^{2}}{2} + \frac{X\Phi'F'}{(1 - M^{2})}],\\
k(u) = - \frac{1}{2}J(u),
\end{eqnarray}
while as before $f(u)$, $g(u)$ and $h(u)$ are still given by (7)-(9) since
 Eq.  (5)  does
not change.
It turns out that, in this case, the reduction equation (14) is identically
 satisfied,  which is reminiscent of the hidden symmetry of Ref. \cite{sch},
so recovering the elliptic solutions found therein.

\section{Behaviour of solutions near magnetic axis}

 It is instructive to analyse the properties of all  possible solutions of
(5) and (6) near  the magnetic axis. Focussing on (5) and (6) we
employ a Cartesian system  $(x,y)$ centred on magnetic axis, i.e.
$R=R(0) +x$ and $z=z(0) +y$, and expand the $u$ surfaces in $x$
and  $y$ around the magnetic axis up to second order:
\begin{equation} u -u(0)= ax^{2} + bxy + cy^{2} + \mbox{higher
orders}.\end{equation} Also, we expand the flux functions
contained in (5) and (6) up to first order in $u-u(0)$, i.e.
\begin{equation}
i(u) =i(0) +i^\prime(0) (u-u(0)),\ \ j(u) =j(0) +j^\prime(0) (u-u(0))
\ \ \mbox{etc},
\end{equation}
 and $R^2$ and $R^4$ up to second order in $x$ and $y$.
On the basis of  the zeroth, first and second order equations thus obtained
from (5) and (6) we can
determine the
coefficients  $a$, $b$ and $c$ of $u-u(0)$.

It turns out that $b = 0$ and
 \begin{equation}
 \frac{a}{c} = \frac{1}{2} + [\frac{1}{4} +
\frac{k(0)R^{2}(0)}{\gamma^2}]^{1/2},
 \end{equation}
 where $\gamma=
[i^\prime(0)+ R(0)^2 j^\prime(0) + R(0)^4 k^\prime(0)]/(2\sqrt{2}).$
 For the cases of subsections 2 and 3  Eq.(29),  on account of  (21) and (26)
implying
 $k(0)<0$,  means that the magnetic surfaces near the magnetic axis are
elliptical
with  elongation being  directed toward $R$. For isothermal magnetic surfaces,
however, the elongation can be either parallel to $R$ if $M^2>1/2$ or parallel
to $z$
if $M^2<1/2$ via (12). For $M^2=1/2$ the isothermal magnetic surfaces become
circular.
  Also,  in all three cases  the ellipses    become  circles
for field aligned flows, i.e. the Palumbo's solution is recovered
 \cite{tas2,sch,pal}.  This point was overlooked in Ref. \cite{tas1}.
In addition, the strength of the toroidal magnetic field on the magnetic axis is
proportional to $k(0)R^{2}(0)$, which vanishes for field aligned flows or
Palumbo's solution.

\section{Conclusions}

In conclusion, the reduction of the "hyperelliptic" equilibria with nonaligned
flows introduced in Ref. \cite{tas1} has been demonstrated for several side
conditions
permitting the discovery of whole classes of magnetohydrodynamic equilibria.
All these equilibria have nonzero magnetic field and elliptic magnetic surfaces
on magnetic axis (see Ref. \cite{sch} for a special case). The
present reduction is based on the annihilation of the fourth and fifth power
under the
square root appearing in Eq.  (13). Another possible reduction could be
sought by finding the conditions under which the fifth order polynomial
appearing in the denominator of Eq.  (13) can be factorized with a double zero
leaving a cubic polynomial under the square root. This and more details about
the equilibria demonstrated here is left to future work.

\begin{center}

{\large\bf Acknowledgements}

\end{center}

Part of this work was conducted during a visit of one of the
authors (G.N.T.) to the Max-Planck-Institut f\"{u}r Plasmaphysik,
Garching. The hospitality of that Institute is greatly
appreciated. The present work was performed under the Contract of
Association ERB 5005 CT 99 0100 between the European Atomic Energy
Community and the Hellenic Republic.

\end{document}